\documentclass[aps,prd,floats,preprintnumbers,superscriptaddress]{revtex4}
\usepackage{hyperref}
\usepackage{epsfig}
\usepackage{amsmath}
\usepackage{amssymb}
\usepackage{amsthm}
\usepackage{srcltx}

\begin{document}

\preprint{NUHEP-TH/12-03}
\preprint{UCB-PTH-12/06}
\preprint{IPMU12-0063}

\title{Neutrino Mixing Anarchy: Alive and Kicking}

\author{Andr\'e de Gouv\^ea}
\affiliation{Northwestern University, Department of Physics \& Astronomy, Evanston, IL~60208, USA}

\author{Hitoshi Murayama}
\affiliation{University of California, Department of Physics, Berkeley, CA~94720, USA}
\affiliation{Lawrence Berkeley National Laboratory, Theoretical Physics Group, Berkeley, CA~94720, USA}
\affiliation{Kavli Institute of Physic and Mathematics of the Universe, University of Tokyo, Kashiwa 277-8583, Japan}


\begin{abstract}

Neutrino mixing anarchy is the hypothesis that the leptonic mixing matrix can be described as the result of a random draw from an unbiased distribution of unitary three-by-three matrices. In light of the recent very strong evidence for a nonzero $\sin^22\theta_{13}$, we show that the anarchy hypothesis is consistent with the choice made by the Nature -- the probability of a more unusual choice is 44\%.  We revisit anarchy's ability to make predictions, concentrating on correlations -- or lack thereof -- among the different neutrino mixing parameters, especially $\sin^2\theta_{13}$ and $\sin^2\theta_{23}$. We also comment on anarchical expectations regarding the magnitude of CP-violation in the lepton sector, and potential connections to underlying flavor models or the landscape.

\end{abstract}

\maketitle


The flavor puzzle has befuddled generations of particle physicists. Since the first years of the quark model and the first successful description of flavor-violating weak processes, the pattern of fermion masses and mixing parameters seems to hint at the existence of some yet-to-be-uncovered organizing principle. The main idea is that new hidden symmetries -- global or local, spontaneously or explicitly broken --  ``explain'' the fact that the charged-fermion masses are very hierarchical and that the quark mixing matrix is very close to the identity matrix.   

The discovery of nonzero neutrino masses and lepton mixing in the end of the last century added new pieces to the flavor puzzle. In particular, the structure of $U$, the leptonic mixing matrix \footnote{Here we assume no new physics beyond masses for the three active neutrinos and mixing among the lepton generations. The neutrino mass eigenstates, with masses $m_{1,2,3}$ are referred to as $\nu_{1,2,3}$, while the neutrino flavor eigenstates are $\nu_{e,\mu,\tau}$.  $U$ can be identified as the matrix that relates these two neutrino bases: $\nu_{\alpha}=U_{\alpha i}\nu_i$, where $\alpha=e,\mu,\tau$, $i=1,2,3$.}, seems to be providing qualitatively different information. Unlike the quark mixing matrix, $U$ cannot be understood as an identity matrix ``perturbed'' by small, hierarchical, off-diagonal matrices. Qualitatively speaking, all elements of the leptonic mixing matrix are large: $|U_{\alpha i}|={\cal O}(1)$.

The flavor literature is densely populated with ingenious attempts to identify the organizing principle behind $U$. The simplest idea, arguably, is to postulate that there is, in fact, no organizing principle behind $U$. Neutrino mixing anarchy \cite{Hall:1999sn} is the hypothesis that $U$ can be described as the result of a random draw from an unbiased distribution of unitary $3\times 3$ matrices, as discussed in detail in \cite{Haba:2000be}. Several years ago \cite{deGouvea:2003xe}, we proposed that a Kolmogorov-Smirnov (KS) test could be employed in order to test the anarchy hypothesis. At the time, anarchy provided a very good fit to the neutrino oscillation data and we further proposed to use the KS statistic in order to make predictions concerning yet-to-be-determined mixing parameters assuming that the anarchy hypothesis is correct. In particular we predicted that, at the two sigma level, $|U_{e3}|^2>0.011$. Here we revisit the anarchy hypothesis in light of new experimental developments regarding the leptonic mixing matrix.





We parameterize $U$ as, 
\begin{equation}
  U = 
  e^{i\eta} e^{i\phi_{1} \lambda_{3} + i \phi_{2} \lambda_{8}}
  \left( \begin{array}{ccc}
  	1 & 0 & 0\\
	0 & c_{23} & s_{23}\\
	0 & -s_{23} & c_{23}
  \end{array} \right)
  \left( \begin{array}{ccc}
  	c_{13} & 0 & s_{13} e^{-i\delta}\\
	0 & 1 & 0\\
	-s_{13}e^{i\delta} & 0 & c_{13}
  \end{array}\right)
  \left( \begin{array}{ccc}
  	c_{12} & s_{12} & 0\\
	-s_{12} & c_{12} & 0\\
	0 & 0 & 1 \end{array} \right)
  e^{i\chi_{1} \lambda_{3} + i \chi_{2} \lambda_{8}},
  \label{eq:U}
\end{equation}
where $\lambda_{3} = {\rm diag}(1, -1, 0)$ and $\lambda_{8} = {\rm diag} (1, 1, -2)/\sqrt{3}$ are Gell-Mann matrices, and $s_{ij} = \sin\theta_{ij}$, $c_{ij}=\cos\theta_{ij}$, $ij=12,13,23$. $\phi_{1,2}$ are unphysical phases that can be absorbed by rephasing the charged lepton fields, $\chi_{1,2}$ are potentially physical ``Majorana'' phases, and $\delta$ is the ``Dirac'' CP-odd phase that is potentially observable in long-baseline neutrino oscillation experiments.

In 2011, data from T2K \cite{Abe:2011sj}, MINOS \cite{Adamson:2011qu}, and Double Chooz \cite{Abe:2011fz} were consistent with $|U_{e3}|^2=\sin^2\theta_{13}\sim 0.02$, and ruled out, for the first time, $|U_{e3}|^2=0$ at the three sigma level. These data were in agreement with previous hints, from combinations of different neutrino experiments, that $|U_{e3}|^2$ did not vanish \cite{Fogli:2008jx}. Very recently, Daya Bay released the analysis of their first data and claimed five sigma evidence that $|U_{e3}|^2$ is not zero \cite{An:2012eh}. Its data point to $|U_{e3}|^2=0.023\pm 0.004$, quite consistent with all previous hints for a nonzero $|U_{e3}|^2$. A few weeks after  \cite{An:2012eh}, RENO released its first results \cite{RENO}, which are in agreement with those from Daya Bay and rule out $|U_{e3}|^2=0$ at more than six sigma. The RENO results point to $|U_{e3}|^2=0.026\pm 0.004$. The other two mixing angles that parameterize $U$ are relatively well-measured from the world's neutrino data (for recent global analyses, see \cite{Fogli:2011qn,Schwetz:2011zk}), and both are large.

 The current data can be summarized as \cite{Schwetz:2011zk,An:2012eh}
\begin{equation}
\sin^2\theta_{12}=0.312^{+0.017}_{-0.015},~~~\sin^2\theta_{23}=0.52\pm0.06,~~~\sin^2\theta_{13}=0.023\pm0.004,
\label{angles}
\end{equation}  
where the last measurement is from Daya Bay. Henceforth, instead of attempting to combine all measuements of $\sin^2\theta_{13}$, we will use the Daya Bay result when it comes to the experimental value of $\sin^2\theta_{13}$. This is both a very good approximation to the world neutrino data and, for our purposes, turns out to be a more conservative choice (slightly smaller central value and larger uncertainty). $\delta$ and the potentially physical Majorana phases remain virtually unconstrained. 

The anarchy hypothesis requires the probability measure of the neutrino mixing matrix to be invariant under changes of basis for the three generations.  It immediately leads to a unique probability distribution for the mixing angles. These can be read off from the invariant integration Haar measure \cite{Haba:2000be},
\begin{equation}
	d U = d s_{12}^{2} \wedge d c_{13}^{4} \wedge d s_{23}^{2} \wedge d\delta
	\wedge d\eta \wedge d\phi_{1} \wedge d\phi_{2} \wedge d\chi_{1} \wedge d\chi_{2},
        \label{eq:Haar}
\end{equation}
up to an overall normalization factor. It yields, for example, probability distributions in $\sin^2 2\theta$ that are the same for $\theta_{12}$, $\theta_{13}$, and $\theta_{23}$ \cite{Haba:2000be}.

In \cite{deGouvea:2003xe}, we derived the probability that the anarchy hypothesis is consistent with experimental data as
\begin{equation}
P_3^{KS}=\epsilon_3\left(1-\log\epsilon_3+\frac{1}{2}\log^2\epsilon_3\right),
\label{P3}
\end{equation}
where 
\begin{equation}
\epsilon_3=2\sin^2\theta_{12}\times 2~\mbox{min} (\sin^2\theta_{23}, \cos^2\theta_{23}) \times2\left(1-\cos^4\theta_{13}\right).
\label{epsilon3}
\end{equation}
The expression reflects the fact that the Haar measure Eq.~(\ref{eq:Haar}) is flat in $s_{12}^2$, $s_{23}^2$, and $c_{13}^4$.  Eq.~(\ref{P3}) is a function of the three mixing angles and is computed by marginalizing over all other parameters, the phases $\delta,\chi_{1,2},\phi_{1,2}$ in Eq.~(\ref{eq:U}). In Eq.~(\ref{epsilon3}), `min' indicates that, depending on whether $\theta_{23}$  is in the ``light side'' ($\theta_{23} < \pi/4$) or the ``dark side'' ($\theta_{23} > \pi/4$) \cite{de Gouvea:2000cq} of its parameter space, one should use either $\sin^2\theta_{23}$ or $\cos^2\theta_{23}$, respectively.  

For the central values of the mixing parameters in Eq.~(\ref{angles}), $\epsilon_3=0.054$ and $P_3^{KS}=44\%$. $P_3^{KS}$ can be interpreted as the probability that a random draw for the mixing matrix will yield a result that is more ``unusual'' than the one made by Nature. In a nutshell, the anarchy hypothesis is consistent with our current understanding of lepton mixing, and the observed values for the mixing parameters are quite typical of the ones obtained by randomly drawing a mixing matrix from an unbiased distribution of unitary $3\times 3$ matrices. 

\begin{figure}[ht]
\includegraphics[width=0.45\textwidth]{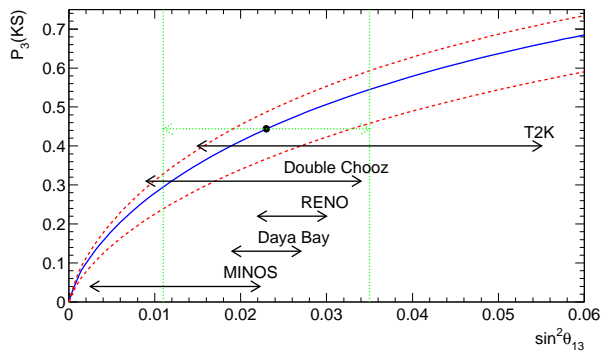}
\caption{$P_3^{KS}$ as a function of $\sin^2\theta_{13}$, for values of $\sin^2\theta_{12,23}$ that independently span the respective allowed regions at the three-sigma confidence level \cite{Schwetz:2011zk}. The solid [blue] line corresponds to the best-fit values, while the dashed [red] lines define the largest and smallest allowed values of  $P_3^{KS}$. The [green] dotted vertical lines indicate the values of $\sin^2\theta_{13}$ allowed by Daya Bay at the three-sigma confidence level. The best estimate for $(\sin^2\theta_{13},P_3^{KS})=(0.023,44\%)$ is indicated by the black dot. The horizontal line segments indicate different experimental measurements of $\sin^2\theta_{13}$ (one-sigma error bars). In the case of MINOS and T2K, we assume a normal neutrino mass hierarchy.}
\label{3dplot}
\end{figure}

Fig.~\ref{3dplot} depicts $P_3^{KS}$ as a function of $\sin^2\theta_{13}$, for values of $\sin^2\theta_{12,23}$ that independently  span their respective allowed regions at the three-sigma confidence level \cite{Schwetz:2011zk}. The dotted vertical lines indicate the values of $\sin^2\theta_{13}$ allowed by Daya Bay at the three-sigma confidence level and allow one to conclude that, within the three-sigma experimentally allowed region of the parameter space, $P_3^{\rm KS}>24\%$.   

\begin{figure}[ht]
\includegraphics[width=0.45\textwidth]{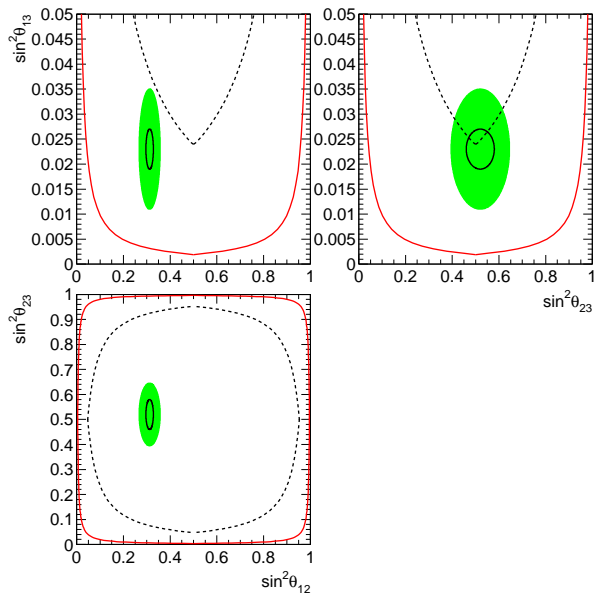}
\caption{Constant $P_2^{KS}(\theta_{ij,ik})=68.3\%$ (``one sigma,'' [black] dashed contours) and $P_2^{KS}(\theta_{ij,ik})=95.5\%$ (``two sigma'' [red] solid contours) contours, in the three distinct $\sin^2\theta_{ij}\times\sin^2\theta_{jk}$ planes. The shaded regions indicate the currently allowed region of the parameter space at the three sigma and one sigma (black ellipses) confidence levels, according to \cite{An:2012eh,Schwetz:2011zk}. Correlations among the experimentally allowed values of different $\sin^2\theta_{ij}$ were not taken into consideration.}
\label{2dplots}
\end{figure}

In order to see how the measurements of the three mixing angles contribute to such an excellent agreement with the anarchy hypothesis, we can define a KS probability $P_2^{KS}(\theta_{ij},\theta_{jk})$, $i,j,k=1,2,3$ as a function of two mixing angles by marginalizing over the ``other'' mixing angle. Following the procedure discussed in detail in \cite{deGouvea:2003xe},
\begin{equation}
P_2^{KS}(\theta_{ij,jk})=\epsilon_2\left(1-\log{\epsilon_2}\right),
\end{equation}
where
\begin{eqnarray}
\epsilon_2(\theta_{12},\theta_{23})&=& 2\sin^2\theta_{12}\times 2~\mbox{min} (\sin^2\theta_{23}, \cos^2\theta_{23}) , \\
\epsilon_2(\theta_{12},\theta_{13})&=&2\sin^2\theta_{12}\times 2\left(1-\cos^4\theta_{13}\right), \\
\epsilon_2(\theta_{13},\theta_{23})&=&2~\mbox{min} (\sin^2\theta_{23}, \cos^2\theta_{23})\times 2\left(1-\cos^4\theta_{13}\right).
\end{eqnarray}
Fig.~\ref{2dplots} depicts contours of constant $P_2^{KS}(\theta_{ij,ik})=68.3\%$ (``one sigma'') and $P_2^{KS}(\theta_{ij,ik})=95.5\%$ (``two sigma'') in the $\sin^2\theta_{ij}\times\sin^2\theta_{jk}$ plane, for the different combinations of $i,j,k$.  The plots also depict the best fits to the current neutrino data at the one and three sigma confidence levels \cite{An:2012eh,Schwetz:2011zk}. Note that we assume that the data responsible for the allowed values of $\sin^2\theta_{12}$, $\sin^2\theta_{13}$, and $\sin^2\theta_{23}$ are completely uncorrelated. While this is not completely correct, it is a good enough approximation for the intentions of this manuscript. The anarchical predictions for the pair-wise values of $\sin^2\theta$ are in good agreement with experimental data (the one-sigma ``tension'' in the $\sin^2\theta_{12}\times\sin^2\theta_{13}$ plane is clearly not statistically significant). This adds credence to the already advertised fact that the anarchy hypothesis is in very good agreement with the neutrino oscillation data. 


Our results do not, of course, imply that anarchy is necessarily correct. Strictly speaking, they simply mean that the neutrino data do not falsify the anarchy hypothesis. Flavor models also make predictions for the values of the mixing angles. We will not attempt to summarize all different possibilities identified in the literature, but will describe in some detail one concrete example, for illustrative purposes.

\begin{figure}[ht]
\includegraphics[width=0.45\textwidth]{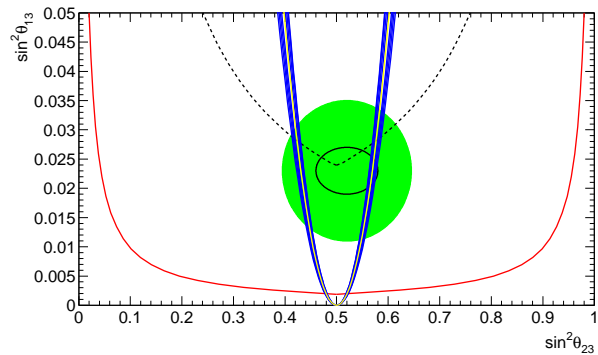}
\caption{Prediction of the ordered hypothesis in the $\sin^2\theta_{23}\times\sin^2\theta_{13}$ plane ([blue] parabolic contours), dictated by Eq.~(\ref{eq:model}) for $C\in[0.8,1.2]$. The light [yellow] curve corresponds to $C=1$. The currently allowed region of the parameter space and the expectations from the anarchy hypothesis, both in Fig.~\ref{2dplots}(top-right), are also depicted.}
\label{with_model}
\end{figure}

Attempts to identify order in the leptonic mixing matrix follow different paths. One is to postulate that a new organizing principle, at leading order, leads to (i) $|U_{\tau i}|=|U_{\mu i}|$, ``maximal'' atmospheric mixing, and (ii) $|U_{e3}|=0$. These combined translate into $\sin^2\theta_{23}=1/2$, $\sin^2\theta_{13}=0$. Other zeroth order predictions may also apply, including $|U_{e2}|^2=|U_{\mu2}|^2=|U_{\tau2}|^2=1/3$, $m_1=m_2=0$, etc. Higher order effects will lead to deviations from these zeroth order predictions. These effects -- their magnitudes and associated flavor structure -- are, of course, model dependent. If, however, the higher order corrections are generic, {\it e.g.}\/, flavor blind and governed by a single small parameter, one generically predicts that deviations of atmospheric mixing from maximal (which can be parameterized by the deviation of $\cos2\theta_{23}$ from zero) are correlated with deviations of  $\sin^2\theta_{13}$ from zero. This is discussed in some detail in, {\it e.g.}\/,~\cite{deGouvea:2004gr}. More quantitatively, one predicts, under the circumstances outlined above,
\begin{equation}
\sin^2\theta_{13}= C\cos^22\theta_{23}=C(1-2\sin^2\theta_{23})^2,
\label{eq:model}
\end{equation}
where $C={\cal O}(1)$, a proportionality constant that is either a free parameter or is specified by the model in question (for a concrete, recent example, see \cite{deAdelhartToorop:2011re}). This relation in the $\sin^2\theta_{23}\times\sin^2\theta_{13}$ plane is depicted in Fig.~\ref{with_model} for $C\in[0.8,1.2]$ (parabolic [blue] region). We refer to this region of the parameter space as the prediction of the `ordered hypothesis.' The figure also depicts the experimentally allowed values  of $\sin^2\theta_{23},\sin^2\theta_{13}$ at the one and three sigma levels, and the region of the parameter space preferred by anarchy at the one and two sigma levels, as in Fig.~\ref{2dplots}(top-right). 

Fig.~\ref{with_model} reveals that the ordered hypothesis and the anarchy one prefer somewhat different regions of the currently allowed $\sin^2\theta_{23}\times\sin^2\theta_{13}$ parameter space. The anarchy hypothesis does not strongly prefer any region of the experimentally available space. It does, however, favor maximal $\sin^2\theta_{23}=1/2$ and ``large'' values of $\sin^2\theta_{13}$. On the other hand, the ordered hypothesis, in light of the Day Bay result, rules out $\sin^2\theta_{23}=1/2$, instead preferring $\cos2\theta_{23}\sim \pm0.1$ (this point was recently also emphasized in \cite{Eby:2011aa}). It is also curious to note that $C\lesssim 0.5$ values are disfavored.

Precision measurements of the neutrino oscillation parameters may ultimately favor anarchy versus order, or vice-versa. The values of the parameters are such that an improved determination of $\sin^2\theta_{23}$ will provide the most discriminating power. If one interprets the width of the blue region in Fig.~\ref{with_model} as indicative of the uncertainty in the ordered predictions, next-generation experiments sensitive to $\delta(\sin^2\theta_{23})\sim 0.02$ -- an uncertainty of a few percent -- would be required to qualitatively change our understanding of structure in lepton mixing. The NO$\nu$A experiment, for example, is aiming at measuring, from $\nu_{\mu}$ disappearance, $\sin^22\theta_{23}$ at the 0.4\% level for $\sin^22\theta_{23}=1$ \cite{Ayres:2004js}, which translates into an uncertainty of 0.03 for $\sin^2\theta_{23}=0.5$. Similar, albeit slightly worse, precision is expected from T2K \cite{Itow:2001ee}. The fact that $\theta_{13}$ is large implies that $\nu_{\mu}\to\nu_e$ searches at T2K and NO$\nu$A, combined with reactor measurements of $\bar{\nu}_e$ disappearance, will allow one to directly measure $\sin^2\theta_{23}$. The precision with which $\sin^2\theta_{23}$ can be measured will be dominated by the precision with which T2K and NO$\nu$A can measure $\sin^2\theta_{23}\sin^22\theta_{13}$,\footnote{$P_{\mu e}\propto\sin^2\theta_{23}\sin^22\theta_{13}$ while $P_{ee}\propto\sin^22\theta_{13}$, to the leading order approximation. The measurement of both observables allows one to disentangle $\sin^22\theta_{13}$ and $\sin^2\theta_{23}$. This, in turn, would allow one to determine whether $\sin^2\theta_{23}$ is smaller or larger than 1/2.} which is expected to be markedly worse than the one advertised for $\sin^2\theta_{23}$ from $\nu_{\mu}$ disappearance, above. Interesting information is also expected from precision measurements of the atmospheric neutrinos at, for example, the INO experiment (see, for example, \cite{Choubey:2005zy,Samanta:2010xm}, and references therein).

Similar arguments can be made in the $\sin^2\theta_{12}\times\sin^2\theta_{13}$ and $\sin^2\theta_{12}\times\sin^2\theta_{23}$ planes. The circumstances here, however, are different. $\sin^2\theta_{12}$ is already known at the few percent level. This implies that constraints on successful ordered scenarios are either very stringent and the associated ``predictions'' are very tight ({\it e.g.}\/, $\sin^2\theta_{12}$ may almost uniquely determine the value of $\sin^2\theta_{13}$ and $\sin^2\theta_{23}$) or correlations are either absent or very weak.  In the $\sin^2\theta_{12}\times\sin^2\theta_{23}$ plane, the anarchical prediction works almost ``too well,'' as the currently three-sigma experimentally allowed region is entirely contained deep in the one-sigma anarchy hypothesis prediction. It is quite unlikely that an ordered hypothesis will lead to a significantly better, statistically speaking, {\it a posteriori}\/ agreement with the data.  


The next obvious target for neutrino oscillation experiments is the discovery of leptonic CP-invariance violation, whose magnitude is governed by the Dirac phase $\delta$. For example, for neutrinos propagating in vacuum, $P(\nu_{\mu}\to\nu_e)-P(\bar{\nu}_{\mu}\to\bar{\nu}_e)\propto\sin\delta$.  Since the Haar measure Eq.~(\ref{eq:Haar}) is flat in $\delta$, the probability distribution of $\sin \delta$ is peaked at $\sin \delta = \pm 1$ \cite{Haba:2000be}: the anarchy hypothesis implies that ``large'' leptonic CP-invariance violation is quite probable.  

If the neutrinos are Majorana fermions, the Majorana phases $\chi_{1,2}$ in Eq.~(\ref{eq:U}) are physical observables.  Similar to that of $\delta$, their probability distributions are flat in $\chi_{1,2}$, respectively.  Majorana phases are known to affect the magnitude of the neutrino exchange contribution to neutrinoless double-beta decay ($0\nu\beta\beta$), and it is interesting to ask whether the anarchy hypothesis has any impact on the expected rates for these rare nuclear processes. The answer, unfortunately, depends on the value of the lightest neutrino mass, which is both experimentally unknown and not addressed by the anarchy hypothesis, which concerns only mixing parameters. Nonetheless, we would like to advertise that, if the anarchy hypothesis is correct and neutrinos are Majorana fermions, it is quite unlikely that the rate for $0\nu\beta\beta$ decay is vanishingly small. 

\begin{figure}[ht]
\includegraphics[width=0.45\textwidth]{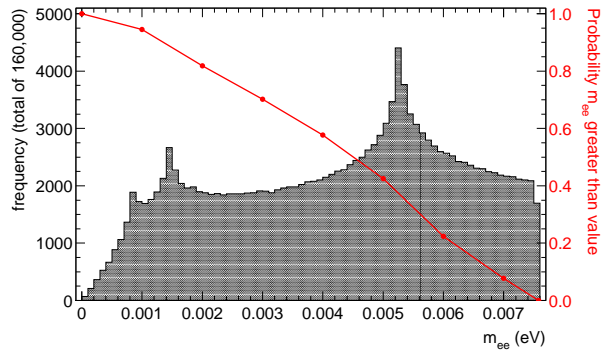}
\caption{Distribution of 160,000 $|m_{ee}|$ values (see Eq.~(\ref{mee})) obtained by scanning over the values of the CP-odd phases $\delta,\chi_{1,2}$ assuming their probability distributions are as prescribed by the anarchy hypothesis. We assume a normal neutrino mass hierarchy and $m_1 = 0.005$~eV, while the mixing angles are fixed at their best fit values, Eq.~(\ref{angles}), along with the mass-squared differences \cite{Schwetz:2011zk}. The [red] curve indicates the probability that a given value of $|m_{ee}|$ or larger is obtained.}
\label{fig:mee}
\end{figure}

For light neutrinos, the amplitude for $0\nu\beta\beta$ is proportional to  
\begin{equation}
  m_{ee} = \sum_i U_{ei}^2 m_i = m_1 (e^{i\chi_1 + i\chi_2/\sqrt{3}} c_{12} c_{13})^2
  + m_2 (e^{-i\chi_1 + i\chi_2/\sqrt{3}} c_{13} s_{12})^2
  + m_3 (e^{-i\delta - 2i\chi_2/\sqrt{3}} s_{13})^2,
  \label{mee}
\end{equation}
using Eq.~(\ref{eq:U}). It is well-known that $m_{ee}$ may completely vanish for a normal neutrino mass hierarchy ($m_1<m_2<m_3$) for  $m_1 \approx 0.005$~eV. Fig.~\ref{fig:mee} depicts a histogram of $|m_{ee}|$ values for $m_1=0.005$~eV obtained by varying $\chi_{1,2}$ and $\delta$ according to their anarchical probability distributions. The [red] curve indicates the probability that a given value of $|m_{ee}|$ or larger is obtained. Tiny $|m_{ee}|$ values are allowed, but in the majority of the cases ($94.5\%$) $|m_{ee}|>0.001$~eV.  Even though this level is beyond the reach of current experimental efforts looking for $0\nu\beta\beta$, it may be probed by future multiton-scale experiments (see, {\it e.g.}\/, \cite{Avignone:2007fu} and references therein). 

The anarchy hypothesis, on the other hand, implies no correlation between the Dirac or Majorana phases with the resulting baryon asymmetry in standard leptogenesis \cite{Xiaochuan}.  This implies that, if the anarchy hypothesis is correct, the direct experimental verification of leptogenesis is impossible. Nonetheless, the positive observation of CP-invariance violation in the lepton sector would still serve as a nontrivial plausibility test.



We conclude by speculating on potential underlying theories beneath the anarchy hypothesis.  Some belong to the category of flavor-symmetry models based on simple $U(1)$'s (as discussed, {\it e.g.}\/, in \cite{Haba:2000be}) where the three generations of lepton doublets are assigned identical flavor quantum numbers.  In this case, the random numbers merely reflect our lack of understanding of the detailed predictions of the models, and the KS test discussed here offers a positive verification of the hypothesis of identical quantum numbers for all the lepton doublets.  

Another class of theories is related to the concept of a string-theory landscape, where our Universe is hypothesized to be only one among a vast number of possible Universes, each with different physics laws and parameters. Virtually all discussions of the landscape, especially when it comes to predictions -- or lack thereof -- rely on the ``anthropic principle,'' where our own existence, or more generically that of galaxies, stars, etc, plays a role in determining which Universes observers are allowed to occupy and, hence, observe.  ``Environmental pressure'' is then defined such that physical parameters, while predicted by the ultimate fundamental physics to be random with complicated probability distributions, are forced towards very special ranges of values and the Universe we inhabit, along with its physics laws and the values of the parameters, tends to be ``on the edge'' for intelligent life, or some appropriate facsimile, to exist. The bottom line is that the observable physics parameters take on what often appears to be ridiculously unlikely values in order to allow observers to exist, but the values are expected to be as mediocre as possible. If the values of the lepton mixing parameters are also a consequence of the landscape, the KS test presented here can be viewed as a direct, faithful test of the landscape hypothesis assuming -- and the test seems to corroborate this! -- leptonic mixing parameters have nothing to do with our existence \cite{Xiaochuan}.

\section*{Acknowledgments}

The work of AdG is sponsored in part by the DOE grant \# DE-FG02-91ER40684. The work of HM was supported in part by the U.S. DOE under Contract DE-AC03-76SF00098, in part by the NSF under grant PHY-1002399, the Grant-in-Aid for scientific research (C) 23540289 from Japan Society for Promotion of Science (JSPS), and in part by World Premier International Research Center Initiative (WPI), MEXT, Japan.

 \end{document}